\documentclass[%
 aip,
 amsmath,amssymb,
 reprint,%
]{revtex4-1}

\usepackage{graphicx}
\usepackage{dcolumn}
\usepackage{bm}

\usepackage[T1]{fontenc}
\usepackage{mathptmx}
\usepackage{etoolbox}
\usepackage{mathtools}

\usepackage[hidelinks]{hyperref}

\usepackage{enumitem}

\makeatletter
\def\@email#1#2{%
 \endgroup
 \patchcmd{\titleblock@produce}
  {\frontmatter@RRAPformat}
  {\frontmatter@RRAPformat{\produce@RRAP{*#1\href{mailto:#2}{#2}}}\frontmatter@RRAPformat}
  {}{}
}%
\makeatother
\begin{document}

\preprint{AIP/123-QED}

\title[]{Inter-role reciprocity in evolutionary trust game on square lattices}
\author{Chaoqian Wang}
\affiliation{School of Mathematics and Statistics, Northwestern Polytechnical University, Xi'an 710072, Shaanxi, China}

\author{Wei Zhang}
\affiliation{Social Networks Lab, ETH Zürich, 8092, Zürich, Switzerland~}
\altaffiliation{Current address is School of Mathematics and Statistics, The University of Sydney, Sydney, NSW 2006, Australia}

\author{Xinwei Wang}
\affiliation{Department of Engineering Mechanics, State Key Laboratory of Structural Analysis, Optimization and CAE Software for Industrial Equipment, Dalian University of Technology, Dalian, 116024, China}

\author{Attila Szolnoki}
\affiliation{Institute of Technical Physics and Materials Science, Centre for Energy Research, P.O. Box 49, H-1525 Budapest, Hungary}
\email{CqWang814921147@outlook.com (C.W.); szolnoki.attila@ek-cer.hu (A.S.)}

\date{\today}

\begin{abstract}
Simulating bipartite games, such as the trust game, is not straightforward due to the lack of a natural way to distinguish roles in a single population. The square lattice topology can provide a simple yet elegant solution by alternating trustors and trustees. For even lattice sizes, it creates two disjoint diagonal sub-lattices for strategy learning, while game interactions can take place on the original lattice. This setup ensures a minimal spatial structure that allows interactions across roles and learning within roles. By simulations on this setup, we detect an inter-role spatial reciprocity mechanism, through which trust can emerge. In particular, a moderate return ratio allows investing trustors and trustworthy trustees to form inter-role clusters and thus save trust. If the return is too high, it harms the survival of trustees; if too low, it harms trustors. The proposed simulation framework is also applicable to any bipartite game to uncover potential inter-role spatial mechanisms across various scenarios.
\end{abstract}

\maketitle

\begin{quotation}
Trust behaviors are ubiquitous in human interactions. How does evolution allow trust in social networks? Since the trust game involves two roles, it is not straightforward to study in a single network. Here, we use a diagonally separated square lattice to simulate trust games. We report that trust can emerge on regular networks through an inter-role spatial reciprocity mechanism. Especially, evolution favors trust and trustworthiness most when the trustee returns moderately. Our results, thus, support field observations and reveal that even minimal network structures can support the emergence of trust.
\end{quotation}

\section{Introduction}

\begin{figure*}[!ht]
	\centering
	\includegraphics[width=.7\textwidth]{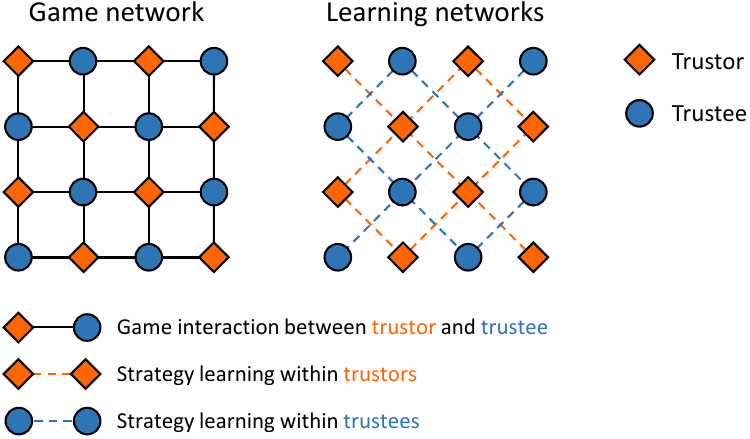}
	\caption{The original square lattice (game network) and its two diagonal sub-lattices (learning networks). Trustors and trustees are arranged in an alternating pattern, allowing them to play games with the opposite role through the horizontal and vertical directions of the original lattice. The two diagonal sub-lattices are disjoint (if $L$ is even), functioning as two independent strategy learning networks where learning only occurs between agents of the same role. }
	\label{fig_demo}
\end{figure*}

Social evolution for various behaviors, such as cooperation~\cite{hamilton1964genetical,axelrod1981evolution}, honesty~\cite{krakauer1995honest,takahashi2012honest,kumar2021evolution}, morality~\cite{boem2012moral,helbing2010moral}, and trust~\cite{abbass2015n,chica2017networked,kumar2020evolution}, reveals how complex social norms and prosocial behaviors can emerge and stabilize in populations through evolutionary mechanisms. In real-world societies, interactions are embedded in specific social network structures, rather than randomly well-mixed populations. This insight has motivated extensive research in spatial evolutionary game theory~\cite{nowak1992evolutionary}, where the network can play a crucial role in the emergence and stability of prosocial behaviors~\cite{allen2017evolutionary}. The research of spatial evolutionary game theory has achieved substantial progress~\cite{perc2017statistical}, both through computer simulations~\cite{szabo1998evolutionary,perc2010coevolutionary} and theoretical analyses~\cite{ohtsuki2006simple,szabo2007evolutionary,wang2024evolutionary}.

Among various social dilemmas, the trust game~\cite{berg1995trust} has a specific feature because it distinguishes between two asymmetric roles: the trustor and the trustee. To capture the essence of trust relationships, the classic trust game involves a sequential interaction in which the trustor first decides whether and how much to invest, while the trustee subsequently chooses whether and how much to return~\cite{cox2004identify,johnson2011trust}. Unlike symmetric games, such as the prisoner's dilemma, the trust game requires interactions across roles. Understanding the evolutionary dynamics of trust, thus, requires accounting for the role asymmetries, as they shape the conditions for the emergence of trust and trustworthiness~\cite{camerer2011behavioral}.

However, simulating asymmetric games, such as the trust game, in structured populations is not straightforward since distinct roles cannot be intuitively differentiated in a single network structure. To avoid this difficulty, previous studies mainly adopted two indirect approaches.

\begin{enumerate}[label=(\arabic*)]
    \item Roles can flow, as in the three-strategy $N$-player trust game~\cite{abbass2015n,greenwood2016finite,chica2017networked,wang2024evolution}. This approach assumes that roles are embedded within strategies, allowing agents to change roles during strategy learning. As a result, it effectively reduces to a multi-strategy symmetric game. Within this framework, previous studies have explored the effects of various additional mechanisms on the evolution of trust, such as a margin system~\cite{guo2024evolution}, loss assessment~\cite{liu2024evolution}, different updating rules~\cite{chica2019effects}, logit dynamics~\cite{towers2017n}, conditional investment in repeated interactions~\cite{liu2022conditional}, diverse investment patterns~\cite{shang2023evolutionary}, reputations~\cite{hu2021adaptive,li2022n,xia2022costly,feng2023evolutionary}, and punishment~\&~reward mechanisms~\cite{chica2019evolutionary,fang2021evolutionary,sun2022evolution,liu2024evolutionary,liu2025evolution,wang2025dynamic}.
    
    \item An individual plays both roles~\cite{burks2003playing}. In this case, the strategies for each role are combined to define four ``composite strategies,'' which compete within the population as a multi-strategy symmetric game. Under this framework, an individual's role remains fixed in strategy learning. Kumar~et~al.\cite{kumar2020evolution} simulated the evolution of trust under this setting in structured populations and found that trust cannot emerge on lattice networks, but can emerge on scale-free networks. Lim and Capraro~\cite{lim2021synergy} and Lim and Masuda~\cite{lim2024truster} also provided theoretical analyses of trust dynamics under this approach.
\end{enumerate}

These two approaches bypass the challenges of simulating multipartite games, but in doing so, they also compromise some essential characteristics of the bipartite scenarios. For instance, in approach~(1), one may question the rationality of the additional assumption---why should a trustee be able to change their social status and become an investor merely because investors achieve higher payoffs, or vice versa? Similarly, in approach~(2), it is unnatural that strategies across different roles are bundled. Even if we accept that an individual could simultaneously act as a trustor and a trustee, it remains unclear why strategies for both roles should change in lockstep. Actually, in a heterogeneous or hierarchical society, these roles cannot be mixed---some individuals are always in a trustor position, while others are mostly in a trustee position. To properly distinguish between different roles in structured populations, more sophisticated frameworks, such as interdependent multilayer networks~\cite{wang2015evolutionary,lugo2015learning}, may be required. In fact, it is because strict role separation complicates simulations in structured populations that the mentioned two approaches have often been adopted as compromises.

Recently, Hauert and Szab{\'o}~\cite{hauert2025phase} pointed out that a square lattice can be divided into two sub-lattices by considering nearest- and next-nearest neighbors separately. If the lattice size is even, then these two sub-lattices are disjoint. On this structure, they observed spontaneous symmetry breaking in symmetric games. For our purposes, this network structure offers a natural solution to the topology required for bipartite games if we, respectively, place trustors and trustees on the two sub-lattices. In this way, the structure reflects two core principles:
\begin{enumerate}[label=(\arabic*)]
    \item Game interactions occur across different roles.
    \item Strategy learning takes place within the community formed by the same role.
\end{enumerate}
These two principles have also been recognized as reasonable modeling choices in previous studies of trust games (for well-mixed populations)~\cite{masuda2012coevolution,tarnita2015fairness,lim2020stochastic,lim2023trust,chiong2022evolution,liu2023n,guo2023evolution,zhou2025evolutionary,su2025time}.

To explore the potential consequence of spatial setting, here, we consider the classic bipartite trust game on a square lattice consisting of two diagonal sub-lattices. The original lattice is used for game interactions across roles, while the two diagonal sub-lattices serve the playground of strategy learning within roles. We find that under a moderate return ratio, this lattice configuration can activate an inter-role spatial reciprocity mechanism to promote the emergence of trust. On this basis, we also explore the relationship between the return ratio $r$ and the investment multiplication factor $u$, detecting an empirical rule $r>1/u$, a critical threshold for the emergence of trust on the proposed topology.

\section{Model}

A trust game is played between two different roles: a trustor and a trustee. A trustor has normalized funds 1 and has two strategies: to invest or not invest. If the trustor does not invest, the player keeps the whole funds 1 as the payoff and the trustee's payoff has to be 0. The investing strategy means giving an $x$ ($0< x\leq 1$) ratio of the funds to the trustee, and the trustor keeps the remaining $1-x$. The investing funds $x$ will be enhanced by a multiplication factor $u$ ($u>1$) so that the trustee can receive $ux$; at this point, the payoffs for both sides further depend on the trustee's strategy: to return or not return. If the trustee does not return (we call such a trustee untrustworthy), the trustor's payoff keeps $1-x$ and the trustee's payoff is $ux$. The return strategy (where the trustee is called trustworthy) means returning a ratio $r$ ($0< r\leq 1$) of the enhanced funds to the trustor and keeps the remaining $1-r$ ratio so that the trustor's payoff is $1-x+rux$ and the trustee's payoff is $(1-r)ux$.

Accordingly, when $1-x+rux>1\Leftrightarrow r>1/u$, the trustor's investment yields a positive feedback if the trustee is trustworthy, and the trustworthy trustee also receives a non-zero payoff. In this way, the collective optimal strategy is for the trustor to invest and the trustee to return. However, since $ux>(1-r)ux$ always holds, not returning is a trustee's individual optimal strategy when facing an investing trustor. Consequently, the trustor loses the investment ($1-x<1$) and, thus, not investing becomes a trustor's individual optimal strategy against an untrustworthy trustee. This intuitively illustrates the collapse of trust in a well-mixed population (see Appendix~\ref{sec_wellmix}).

To model a structured population, we consider an $L\times L$ square lattice with periodic boundary conditions. Each lattice site $i$ is occupied by an agent, whose position keeps unchanged in the evolutionary process. The role of agent $i$ is represented by $\text{role}(i)\in\{1,2\}$, where $\text{role}(i)=1$ denotes a trustor and $\text{role}(i)=2$ denotes a trustee. We arrange trustors and trustees in an alternating pattern on the square lattice, as shown in Fig.~\ref{fig_demo} (left). This regular lattice is referred to as the game network, where each agent interacts with their four nearest neighbors as game partners. Let $\Omega_i^+$ denote the set of game neighbors of agent $i$. Clearly, if $\text{role}(i)=1$, then $\text{role}(j)=2$ for $j\in \Omega_i^+$ and vice versa. This ensures that game interactions occur only between a trustor and a trustee.

We define the four next-nearest (diagonal) neighbors of each agent as their learning neighbors, as shown in Fig.~\ref{fig_demo} (right). When the lattice size $L$ is even, this leads to two disjoint sub-lattices, one consisting purely of trustors and the other of trustees. Let $\Omega_i^\times$ denote the set of learning neighbors of agent $i$. As seen in Fig.~\ref{fig_demo} (right), $\text{role}(j)=\text{role}(i)$ for $j\in \Omega_i^\times$ and vice versa. This ensures that strategy learning occurs only between agents of the same role.

The current strategy of agent $i$ is denoted by $\mathbf{s}_i\in\{(1,0),(0,1) \}$, where $\mathbf{s}_i=(1,0)$ represents investing (if agent $i$ is a trustor) or returning (if agent $i$ is a trustee), and $\mathbf{s}_i=(0,1)$ means not investing (if trustor) or not returning (if trustee). For convenience, we use only $(1,0)$ and $(0,1)$ and do not mathematically distinguish the strategy representations for different roles. Based on the earlier description of the trust game, the payoff matrix for a trustor [$\text{role}(i)=1$] can be written as 
\begin{equation}\label{eq_m1}
    \mathbf{M}_1=
    \begin{pmatrix}
    1-x+rux & 1-x\\
    1 & 1
    \end{pmatrix}.
\end{equation}
An investing trustor has $1-x+rux$ if the trustee is trustworthy and $1-x$ otherwise. A non-investing trustor always keeps $1$. Similarly, the payoff  matrix for a trustee [$\text{role}(i)=2$] is given by 
\begin{equation}\label{eq_m2}
    \mathbf{M}_2=
    \begin{pmatrix}
    (1-r)ux & 0\\
    ux & 0
    \end{pmatrix}.
\end{equation}
If the trustor invests, a trustworthy trustee has $(1-r)ux$ and an untrustworthy trustee has $ux$. A trustee always has $0$ if the trustor does not invest.

The payoff $\pi_i$ of agent $i$ is determined by the current strategies of themselves and their game neighbors $l\in\Omega_i^+$. Specifically, agent $i$ plays trust games with all their game neighbors and takes the averaged payoffs obtained from these interactions. Therefore, the payoff $\pi_i$ of agent $i$ is calculated by 
\begin{equation}\label{eq_payoff}
    \pi_i=\frac{1}{|\Omega_i^+|}\sum_{l\in\Omega_i^+}\mathbf{s}_i \cdot\mathbf{M}_{\text{role}(i)} \cdot \mathbf{s}_l^\top,
\end{equation}
where $|\Omega_i^+|\equiv 4$ is the number of game neighbors.

In each elementary Monte Carlo (MC) step, a random agent $i$ is selected to update the strategy. First, the payoff $\pi_i$ is calculated by Eq.~(\ref{eq_payoff}). Then, agent $i$ randomly chooses a learning neighbor $j\in\Omega_i^{\times}$, whose payoff $\pi_j$ is calculated according to the same principle. Agent $i$ adopts the strategy of agent $j$, with a probability $W_{\mathbf{s}_i \gets \mathbf{s}_j}$ given by the pairwise comparison rule~\cite{szabo1998evolutionary},
\begin{equation}\label{eq_pc}
    W_{\mathbf{s}_i \gets \mathbf{s}_j}=
    \frac{1}{1+\exp{(-(\pi_j-\pi_i)/\kappa)}},
\end{equation}
where, $\kappa=0.1$ is a noise parameter. According to Eq.~(\ref{eq_pc}), the greater the advantage of agent $j$'s payoff $\pi_j$ over $\pi_i$, the more likely agent $i$ is to adopt the strategy of agent $j$. A full MC step consists of $L^2$ such elementary steps, and strategies in the system evolve across the population. Since trustors and trustees each account for $L^2/2$ of the population, there is no sampling bias between the different roles.

\section{Results}

\begin{figure*}[!ht]
	\centering
	\includegraphics[width=\textwidth]{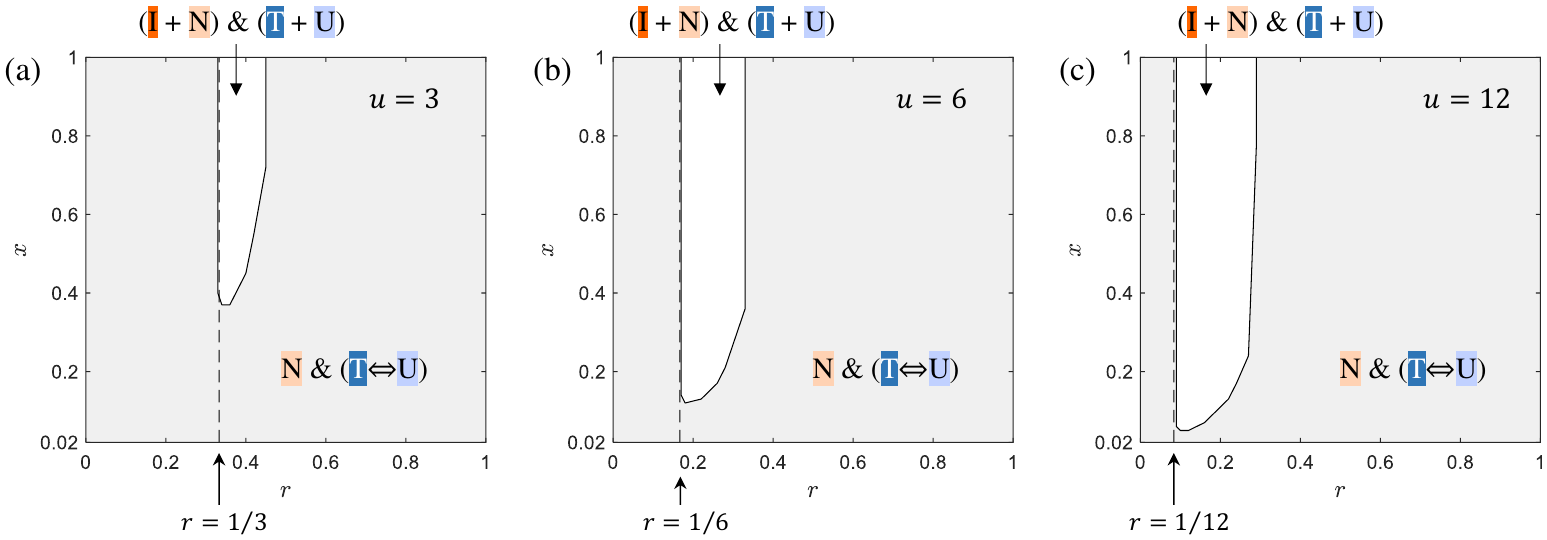}
	\caption{Phase diagrams of the system behavior with respect to return ratio $r$ and investment ratio $x$ for different multiplication factors $u=3,6,12$. Trust emerges in the $(\text{I}+\text{N})\&(\text{T}+\text{U})$ phase (white region) with large $x$ and moderate $r$, where the four strategies of the two roles coexist. Trust cannot survive in the $\text{N}\&(\text{T}\Leftrightarrow\text{U})$ phase (gray region), where trustors do not invest, and trustworthy and untrustworthy trustees are in neutral drift. The dashed line marks $r=1/u$ in each panel.}
	\label{fig_xr}
\end{figure*}

\begin{figure*}[!ht]
	\centering
	\includegraphics[width=\textwidth]{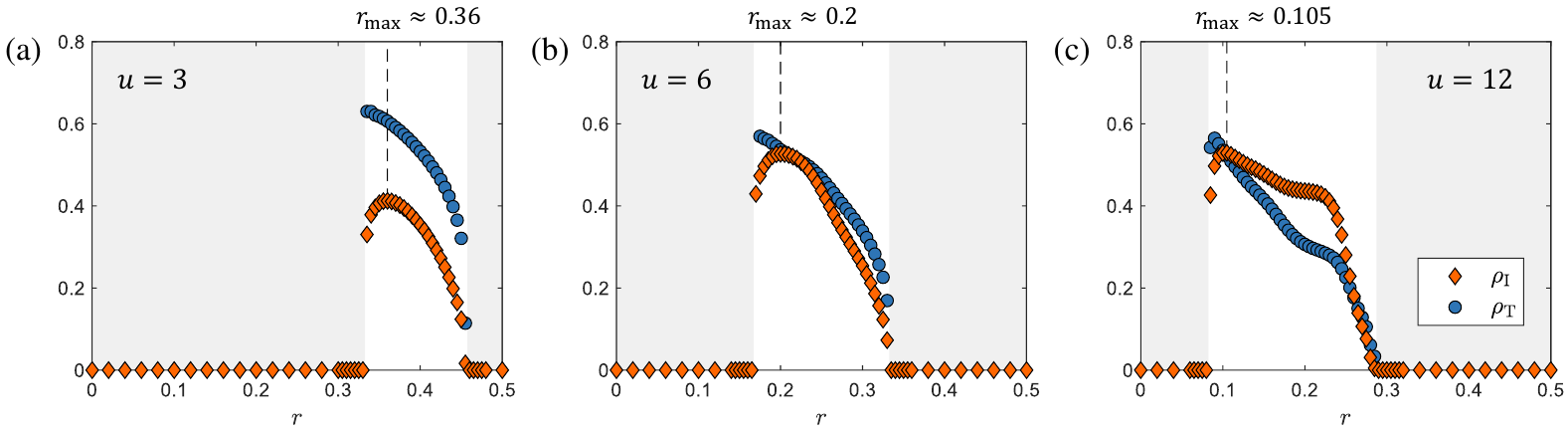}
	\caption{The proportions of investing trustors ($\rho_\text{I}$) and trustworthy trustees ($\rho_\text{T}$) as a function of return ratio $r$ for different multiplication factors $u=3,6,12$. A moderate return ratio maximizes trust, and the optimal return ratio is $r_\text{max}\approx 0.36, 0.2, 0.105$ for $u=3,6$, and $12$, respectively. The gray intervals do not present $\rho_\text{T}$ because it is not stationary in neutral drift. Other parameter: $x=1$.}
	\label{fig_1Dr}
\end{figure*}

\begin{figure*}[!ht]
	\centering
	\includegraphics[width=\textwidth]{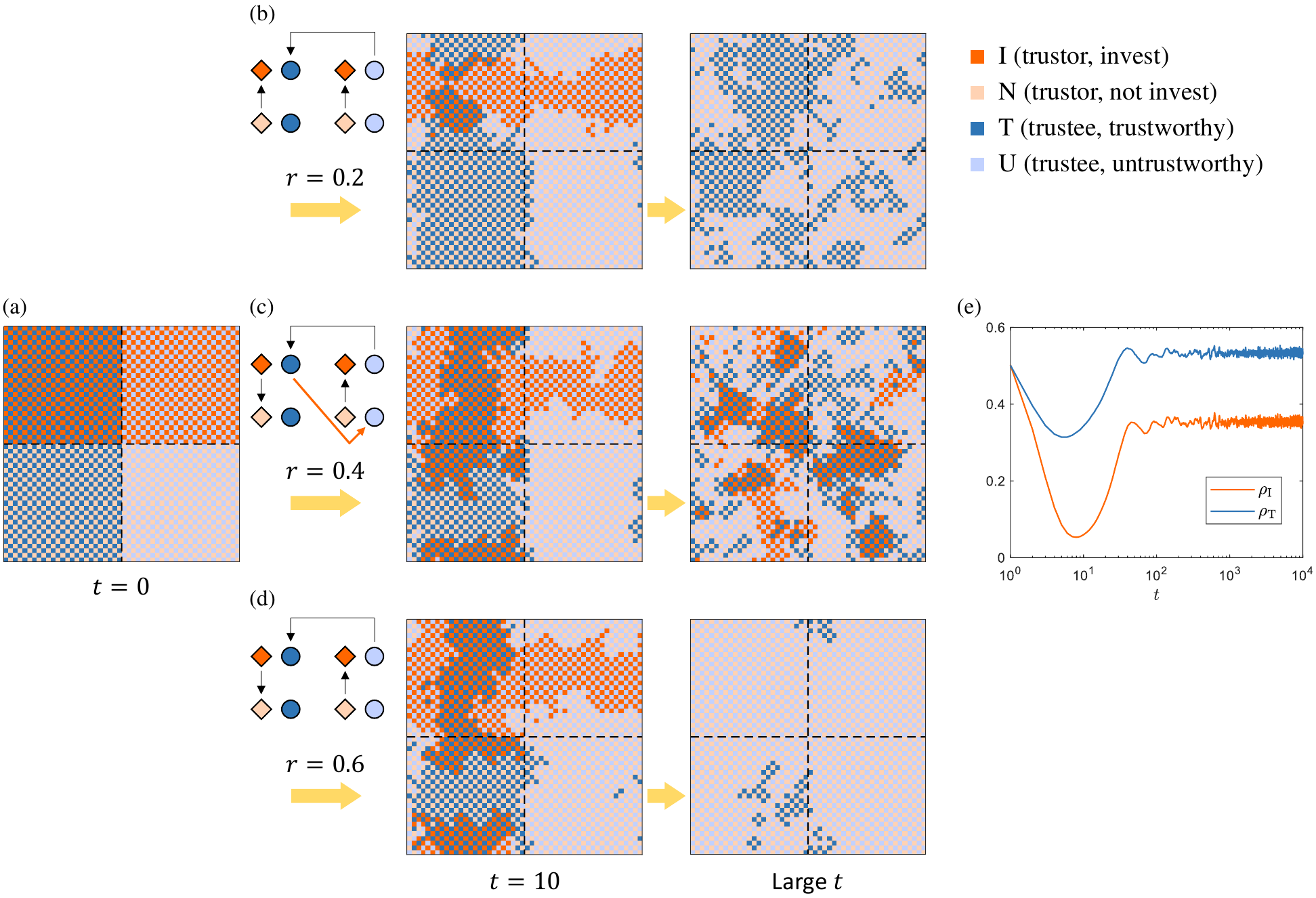}
	\caption{Visualization of trust emergence (at moderate return ratios) and trust extinction (at high and low return ratios) on $50\times 50$ lattices. (a) The initial preparation divides the lattice into four blocks, $\text{I}\&\text{T}$, $\text{I}\&\text{U}$, $\text{N}\&\text{T}$, and $\text{N}\&\text{U}$, to better observe their competition. (b) At low return ratio ($r=0.2$), non-investing trustors always invade investing trustors, and trust cannot survive. (c) At moderate return ratio ($r=0.4$), investing trustors can invade non-investing trustors when faced with trustworthy trustees, and trust can emerge. (d) At high return ratio ($r=0.6$), trustworthy trustees are unable to invade untrustworthy trustees, and trust cannot survive. In (b)--(d), untrustworthy trustees always invade trustworthy trustees when faced with investing trustors. The arrows represent the direction of invasions. (e) At $r=0.4$, the proportions of investing trustors and trustworthy trustees achieve non-equilibrium statistical steady states on $600\times 600$ lattices with a random initial strategy distribution. Other parameters: $u=3$, $x=1$.}
	\label{fig_snap4}
\end{figure*}

\begin{figure*}[!ht]
	\centering
	\includegraphics[width=.9\textwidth]{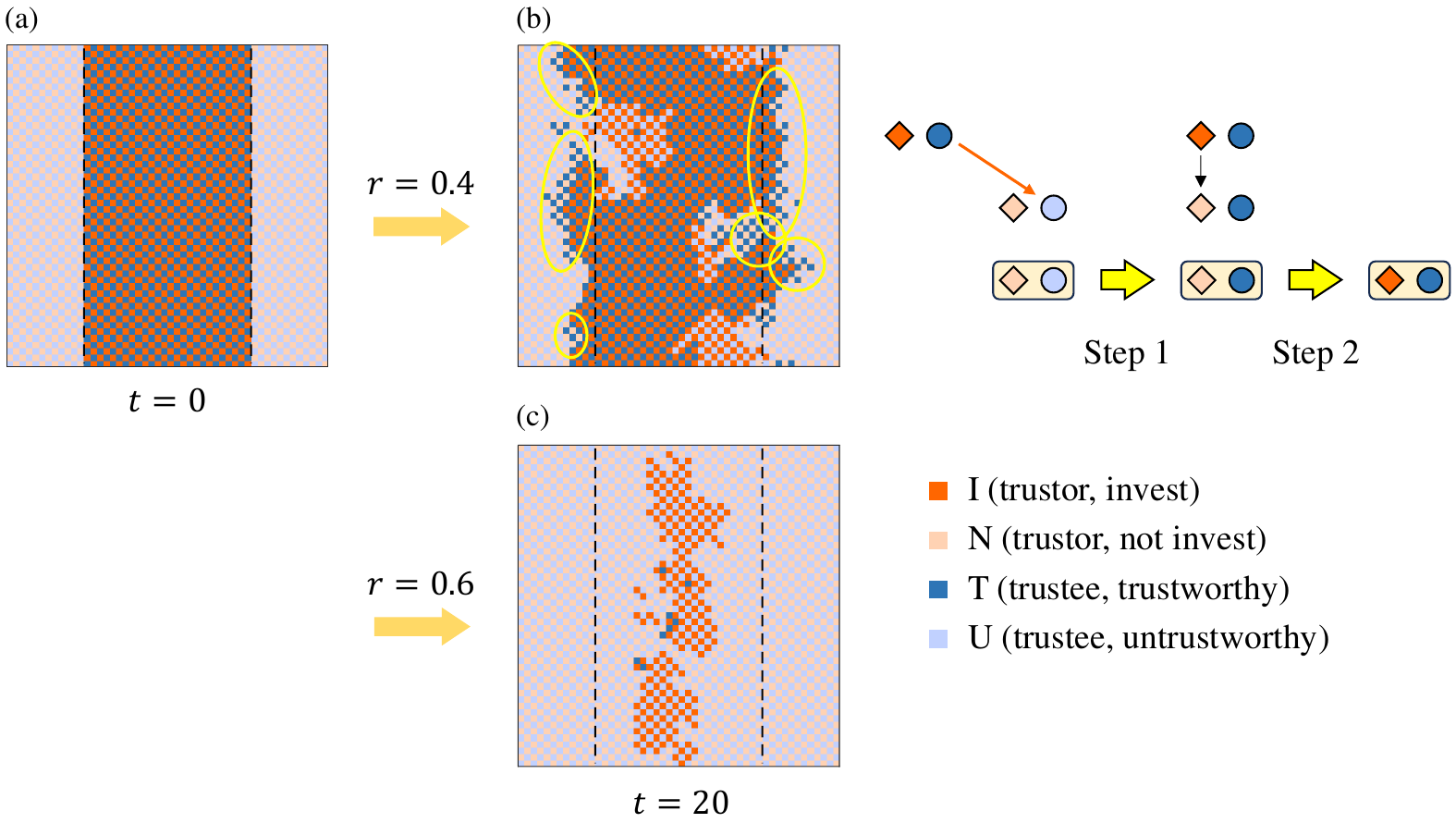}
	\caption{Inter-role spatial reciprocity in the two-step invasion $\text{I}\&\text{T}\rightarrow \text{I}\&\text{U} \rightarrow \text{N}\&\text{U}$ is key to trust emergence. (a) The initial preparation divides the lattice into two blocks, $\text{I}\&\text{T}$ and $\text{N}\&\text{U}$, to better observe the two-step invasion. (b) At moderate return ratio ($r=0.4$), trustworthy trustees are supported by neighboring investing trustors, thus invading untrustworthy trustees near non-investing trustors (Step~1, $\text{I}\&\text{T}\rightarrow \text{I}\&\text{U}$, regions marked by yellow circles); near trustworthy trustees, an investing trustor has higher fitness and invades non-investing trustors (Step~2, $\text{I}\&\text{U} \rightarrow \text{N}\&\text{U}$). (c) At high return ratio ($r=0.6$), trustworthy trustees have low fitness due to returning too much, thus unable to make the invasion in Step~1. Other parameters: $u=3$, $x=1$. }
	\label{fig_snap2}
\end{figure*}

\begin{figure}[!ht]
	\centering
	\includegraphics[width=.5\textwidth]{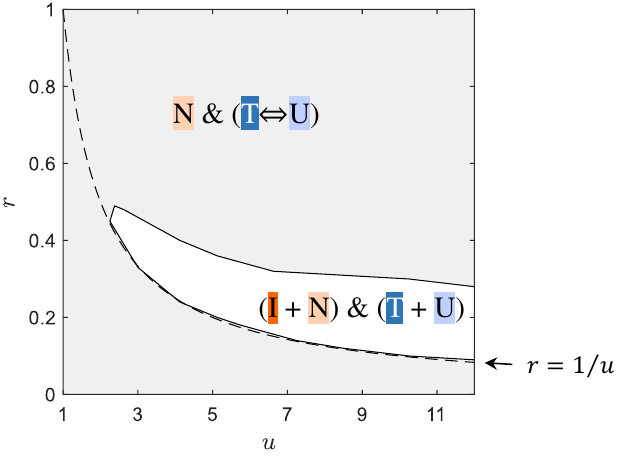}
	\caption{Phase diagram of the system behavior with respect to multiplication factor $u$ and return ratio $r$. Evolution favors trust in the $(\text{I}+\text{N})\&(\text{T}+\text{U})$ phase with moderate $ur$ and does not favor trust in the remaining $\text{N}\&(\text{T}\Leftrightarrow\text{U})$ phase. The system behavior of each phase is the same as in Fig.~\ref{fig_xr}. Intuitively, a larger multiplication factor $u$ means that investing trustors can maintain the same advantage $ur$ with a lower return rate $r$, allowing trustworthy trustees to retain a greater share of the benefits that can be used to compete against untrustworthy trustees. The dashed line marks $r=1/u$. Other parameter: $x=1$.}
	\label{fig_ru}
\end{figure}

\begin{figure*}[!ht]
	\centering
	\includegraphics[width=\textwidth]{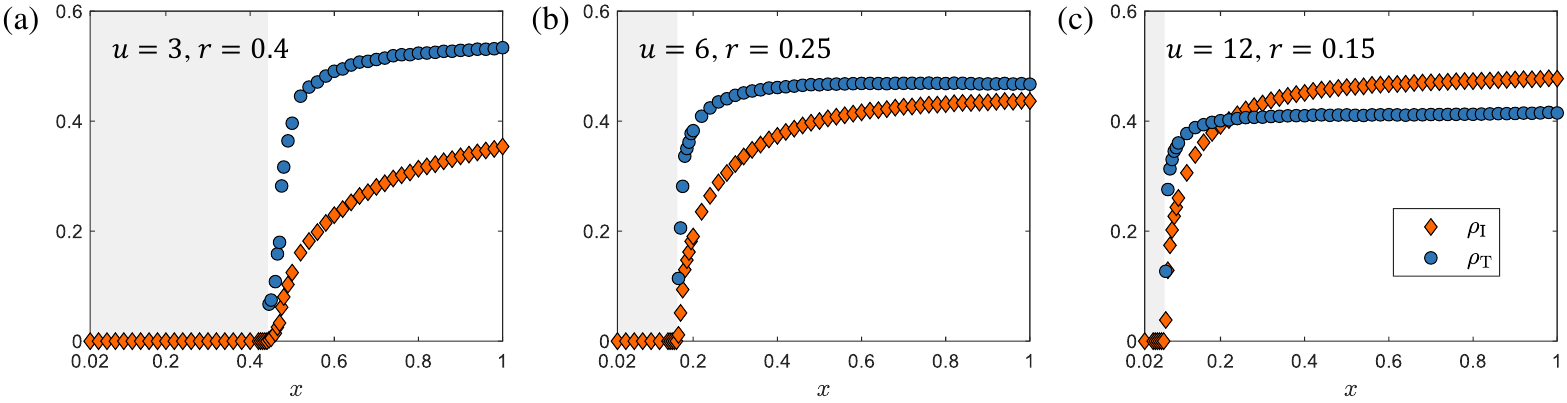}
	\caption{The proportions of investing trustors ($\rho_\text{I}$) and trustworthy trustees ($\rho_\text{T}$) as a function of investing ratio $x$ for different multiplication factors and return ratios $(u,r)=(3,0.4),(6,0.25),(12,0.15)$. A large investing ratio $x$ promotes trust because it brings trustors more return when trust and trustworthiness coexist. $\rho_\text{T}$ within the gray intervals is in neutral drift and not presented. }
	\label{fig_1Dx}
\end{figure*}

As a consequence of our model definition, there are four strategies in the system: I (trustor, Invest), N (trustor, Not invest), T (Trustworthy trustee), and U (Untrustworthy trustee). Their proportions within each role are denoted as $\rho_\text{I}$, $\rho_\text{N}=1-\rho_\text{I}$, $\rho_\text{T}$, and $\rho_\text{U}=1-\rho_\text{T}$. The system state is, thus, represented by $0\leq \rho_\text{I}\leq 1$ and $0\leq \rho_\text{T}\leq 1$.

Unless otherwise specified, all simulations are launched from the configuration shown in Fig.~\ref{fig_snap4}(a). Compared to random initialization, such a prepared initial state, while ensuring fair initial strategy proportions of $\rho_\text{I}, \rho_\text{T}\approx 50\%$, can better avoid the loss of solutions due to early extinction~\cite{szolnoki2011pre,szolnoki2017njp}. We simulate $1\times 10^4$ full MC steps and take the averaged $\rho_\text{I}$ and $\rho_\text{T}$ over the last $5\times 10^3$ full steps as the results, which reflects a non-equilibrium statistical steady state (see Fig.~\ref{fig_snap4}(e) for an illustration) or fixation states. Depending on the required simulation accuracy, we use lattice sizes of $L=300$ or $L=600$ (which are sufficiently large) in main simulations, while a smaller lattice with $L=50$ is used to demonstrate spatial mechanisms (e.g., Figs.~\ref{fig_snap4} and \ref{fig_snap2}).

We find two phases in the non-equilibrium statistical steady or fixation states (Fig.~\ref{fig_xr}). One is $(\text{I}+\text{N})\&(\text{T}+\text{U})$, where both strategies coexist within each role. In this phase, trust emerges and is sustained. Another is $\text{N}\&(\text{T}\Leftrightarrow\text{U})$, where investing trustors go extinct, leaving only non-investing trustors. As a result, the two trustee strategies yield indistinguishable payoffs (0) and are in neutral drift (i.e., the system eventually retains either T or U at random, with probabilities determined by their proportions at I's extinction~\cite{cox1983voter}). We say that trust cannot survive in this phase. According to Figs.~\ref{fig_xr}(a)--(c), the $(\text{I}+\text{N})\&(\text{T}+\text{U})$ phase exists under moderate return ratio 
$r$ and large investing ratio 
$x$. Moreover, a higher multiplication factor 
$u$ expands the parameter region of the $(\text{I}+\text{N})\&(\text{T}+\text{U})$ phase.

The trust emergence phase $(\text{I}+\text{N})\&(\text{T}+\text{U})$ seems to correlate with the following condition:
\begin{equation}
    r>\frac{1}{u}.
\end{equation}
For $u=3,6,12$, the left boundary of the $(\text{I}+\text{N})\&(\text{T}+\text{U})$ phase is $r=1/3,1/6,1/12$, respectively, as indicated by the dashed lines in Figs.~\ref{fig_xr}(a)--(c). This is consistent with theoretical analysis of a single trust game. A trustor can obtain a net benefit from investment only when $1-x+rux>1\Leftrightarrow r>1/u$, and only under this condition can a trustee's return incentivize the trustor to invest. The evolutionary dynamics on square lattices reproduce this local property, hence the empirical condition $r>1/u$ for the emergence of trust.

We further study the proportions of prosocial strategies $\rho_\text{I}$ and $\rho_\text{T}$ as a function of return ratio $r$ under $x=1$ (Fig.~\ref{fig_1Dr}). At this point, investing trustors give all of their funds. We find that as $r$ increases, the proportion of investing trustors initially increases and ultimately decreases; that is, a moderate return ratio is most favorable for trust. For $u=3,6,12$ (i.e., $1/u\approx 0.333,0.167,0.083$), the return ratios that maximize $\rho_\text{I}$ are approximately $r_\mathrm{max}\approx 0.36,0.2,0.105$. The optimal return ratio $r_\mathrm{max}$ for trust roughly satisfies the relation $r_\mathrm{max}\gtrapprox 1/u$. On the other hand, in the $(\text{I}+\text{N})\&(\text{T}+\text{U})$ phase, the proportion of trustworthy trustees $\rho_\text{T}$ peaks near $r\approx 1/u$; beyond this point, their advantage diminishes and their proportion declines. The transitions of $\rho_\text{I}$ and $\rho_\text{T}$ near $r\approx 1/u$ appear to be discontinuous, whereas the subsequent transitions to trust extinction at larger $r$ are continuous.

To intuitively explain why a moderate return ratio favors trust, we demonstrate the evolutionary patterns at $r=0.2,0.4,0.6$ on a $50\times 50$ lattice (Fig.~\ref{fig_snap4}). The initial condition consists of four blocks, $\text{I}\&\text{T}$, $\text{I}\&\text{U}$, $\text{N}\&\text{T}$, and $\text{N}\&\text{U}$, which allows an observation of the competition between different strategies. We first observe two basic facts: (1) The U strategy in the $\text{I}\&\text{U}$ block always invades the T strategy in $\text{I}\&\text{T}$ [Figs.~\ref{fig_snap4}(b)--(d)] since when the trustor invests, a trustee who keeps the entire investment always gains a higher payoff than one who returns a portion. (2) The N strategy in $\text{N}\&\text{U}$ always invades the I strategy in $\text{I}\&\text{U}$, because when trustees return nothing, a trustor who retains their funds always has a higher payoff than one who invests. These are also the core mechanisms behind the collapse of trust. Building on this, different evolutionary patterns emerge under varying return ratios $r$ and attempt to counteract the extinction of trust.

Specifically, extremely low or high return ratios $r$ fail to prevent the extinction of trust (Figs.~\ref{fig_snap4}(b) and (d)). At a low return ratio ($r=0.2$), the N strategy in $\text{N}\&\text{T}$ also invades the I strategy in $\text{I}\&\text{T}$, because the feedback $rux$ that investing trustors receive from trustworthy trustees is too small to provide a competitive advantage against non-investing trustors. As a result, strategy N always invades strategy I in the system regardless of the trustee's behavior, leading to the collapse of trust.

At a high return ratio ($r=0.6$), the I strategy in $\text{I}\&\text{T}$ is able to invade the N strategy in $\text{N}\&\text{T}$, as investing trustors receive a relatively high payoff $rux$ from trustworthy trustees. However, the excessively high return ratio also means that the amount retained by trustworthy trustees, $(1-r)ux$, becomes too small, thereby accelerating the rate at which strategy U invades strategy T. This effect is evidenced by the lower proportion of the T strategy in the large $t$ snapshots in Figs.~\ref{fig_snap4}(b) and (d), where the residual strategy T after trust collapse is much fewer for $r=0.6$ than for $r=0.2$. As a result, the invasion by N against I in $\text{N}\&\text{U}\to\text{I}\&\text{U}$ dominates over the invasion by I against N in $\text{I}\&\text{T}\to\text{N}\&\text{T}$, leading to the extinction of trust.

In contrast, a moderate return ratio ($r=0.4$) gives rise to a mechanism that counteracts the extinction of trust (Fig.~\ref{fig_snap4}(c)). Although the left snapshot ($t=10$) in Fig.~\ref{fig_snap4}(c) appears similar to that in Fig.~\ref{fig_snap4}(d), the moderate return ratio induces a non-negligible strategy flow, as indicated by the orange arrow in the left diagram of Fig.~\ref{fig_snap4}(c), where the T strategy in $\text{I}\&\text{T}$ can invade the U strategy in $\text{N}\&\text{U}$. This process will be discussed in more detail in Fig.~\ref{fig_snap2}. Eventually, at large $t$, a dynamic balance emerges among all four strategies, which allows trust to persist. As additional supporting evidence, Fig.~\ref{fig_snap4}(e) shows the strategy evolution on a $600\times 600$ lattice under random initial conditions at $r=0.4$. It demonstrates a typical spatial evolutionary dynamic, where the proportions of investing trustors and trustworthy trustees first decline and then rise~\cite{perc2008restricted,szolnoki2009promoting}, eventually fluctuating around a non-equilibrium statistical steady state.

We attribute the emergence of trust to an inter-role spatial reciprocity mechanism in the two-step invasion sequence $\text{I}\&\text{T}\rightarrow \text{I}\&\text{U} \rightarrow \text{N}\&\text{U}$. To demonstrate this mechanism, we prepare a $50\times 50$ lattice with the initial condition shown in Fig.~\ref{fig_snap2}(a), which allows an observation of the competition between $\text{I}\&\text{T}$ and $\text{N}\&\text{U}$. Under a moderate return ratio, trust can spread through the invasion sequence $\text{I}\&\text{T}\rightarrow \text{I}\&\text{U} \rightarrow \text{N}\&\text{U}$. In Step~1, the T strategy in $\text{I}\&\text{T}$ receives investment support from neighboring I-players and has a higher payoff than the U strategy in $\text{N}\&\text{U}$. In this way, the T strategy here can invade U and play the role of ``pioneer colonizers,'' as indicated by the yellow elliptical regions in Fig.~\ref{fig_snap2}(b). These pioneering T-players, together with nearby non-investing trustors in the newly expanded area, form $\text{N}\&\text{T}$ pairs. In Step~2, the non-investing trustors in the new $\text{N}\&\text{T}$ regions find themselves receiving lower payoffs than the investing trustors from the direction the pioneering T strategy came from, since a trustor's investment is always rewarded in the presence of trustworthy trustees. Consequently, the I strategy follows the expanding front of the T strategy, enabling the invasion $\text{I}\&\text{U} \rightarrow \text{N}\&\text{U}$ and consolidating the transformation of $\text{N}\&\text{U}$ into $\text{I}\&\text{T}$. In contrast, as shown in Fig.~\ref{fig_snap2}(c), under an excessively high return ratio, the T strategy in $\text{I}\&\text{T}$ returns too much to the trustor and, thus, lacks sufficient payoff advantage to act as a pioneer in invading untrustworthy trustees, making it unable to initiate Step~1 in the two-step invasion.

Furthermore, the specific value of the moderate return ratio $r$ is related to multiplication factor $u$. As shown in Fig.~\ref{fig_ru}, a higher multiplication factor $u$ requires a smaller return ratio $r$ for the evolution of trust. Intuitively, an investing trustor only needs a sufficient return (i.e., $1-x+rux$ in $\mathbf{M}_1$) to outperform a non-investing trustor. In other words, the essence of ``moderate return promotes trust'' lies in keeping $1-x+rux$ at a moderate level. If the multiplication factor $u$ is large, the same value of $1-x+rux$ can be maintained at a lower return ratio $r$. In this case, the trustworthy trustee can retain a larger portion of the investment, $(1-r)ux$, which strengthens their ability to resist untrustworthy trustees and better support the investing trustor. The dashed line $r=1/u$ in Fig.~\ref{fig_ru} aligns well with the phase boundary obtained from agent-based simulations and illustrates, over a broader parameter range, the conjecture proposed in Fig.~\ref{fig_xr} that trust emerges when $r>1/u$.

In the preceding analyses, we fixed $x=1$. We finally investigate the role of the investing ratio $x$. Fig.~\ref{fig_1Dx} shows how increasing $x$ affects the proportions of investing trustors and trustworthy trustees under several trust-emergence $(u,r)$ combinations. We find that increasing the investing ratio $x$ simultaneously promotes trust and trustworthiness. This is because, when $(u,r)$ supports a trust-emergence mechanism, a larger investing ratio yields greater returns within that mechanism, thereby amplifying its effect. Moreover, we observe that when $x$ is sufficiently large, the values of $\rho_\text{I}$ and $\rho_\text{T}$ become relatively insensitive to further changes in $x$. Therefore, setting $x=1$ not only has an intuitive interpretation as ``full investment,'' but also provides a numerically stable choice.

\section{Discussion}
How does the classic bipartite trust game evolve in structured populations? To address this question, we applied a topology recently suggested by Hauert and Szab{\'o}~\cite{hauert2025phase}, dividing a square lattice into two diagonal sub-lattices for strategy learning and keeping the original lattice for game interactions. This topology appears to provide a minimal spatial structure that enables interactions across roles and strategy learning within roles. Such role differentiation is essential to reflect the characteristics of the classic trust game.

Our simulation revealed an inter-role spatial reciprocity mechanism, through which trust and trustworthiness can emerge on square lattices. Specifically, a moderate return ratio is optimal for the evolution of trust. Under such conditions, the inter-role spatial reciprocity mechanism is activated in the trust game, allowing trust and trustworthiness to expand through a two-step invasion process. First, trustworthy trustees, supported by neighboring investing trustors, invade untrustworthy trustees adjacent to non-investing trustors, creating a positive environment for investment. Second, investing trustors follow in the footsteps of these pioneering trustworthy trustees, completing the expansion of trust. This spatial pattern, with trustworthiness leading and trust closely following, is similar to the ``coordinated growth'' in previous the literature~\cite{szolnoki2013information,lugo2015learning,wang2023public}.

In an earlier study, Kumar~et~al.~\cite{kumar2020evolution} investigated a four-strategy (combination of two strategies for two roles) trust game model on a square lattice. Their results indicated that trust cannot survive on regular networks, such as square lattices. In contrast, we studied the original bipartite trust game. Our results illustrated that, when following the role separation assumption in the classic trust game, trust can indeed emerge on a square lattice.

Regarding the payoff structure, we followed the same parameterization used by Kumar~et~al.~\cite{kumar2020evolution}, who respected the original experimental trust game~\cite{berg1995trust}, which assumed a multiplication factor $u=3$ for investment. However, we further explored the effect of the multiplication factor $u$ and revealed an empirical $r>1/u$ rule---trust emerges on lattice networks when the return ratio $r$ exceeds $1/u$. Intuitively, this threshold implies that the trustee's return can provide positive feedback to the trustor's investment in a local trust game. We also provided intuitive explanations for why overly small or large return ratios hinder the emergence of trust---small returns disadvantage the trustor, while large returns disadvantage the trustee. These local properties are faithfully reproduced on the lattice network and its sub-lattice structures.

It is worth noting that the lattice setup in this work is fortunate and incidental, which comes with several limitations. For example, it only supports the four-neighbor setting. Slight variations, such as the eight-neighbor setup, cannot be intuitively implemented through diagonal sub-lattices. Moreover, it can only distinguish two roles. Models involving three or more roles cannot be directly realized on our lattice. In essence, asymmetric games simply require a generalized network structure that supports cross-role interactions and within-role strategy learning. Another parallel implementation to this property is multilayer networks~\cite{lugo2015learning,hauert2024spontaneous,ling2025supervised}. However, we would like to emphasize that the topology setup in our work offers an especially minimal construction that enables elegant spatial visualization, which is also applicable to any bipartite game. The spatial mechanism of other bipartite games in structured populations, such as the ultimatum game~\cite{guth1982experimental,page2000spatial} and the Stackelberg game~\cite{stackelberg1934marktform}, can also be investigated on the same topology. 

Building on the classic trust game, one can explore how heterogeneity in parameters~\cite{perc2008hetero,wang2022involution} (e.g., $x,r,u$) affects the emergence of trust. Furthermore, while our lattice topology limits the number of roles, it does not constrain the number of strategies. In other words, it allows for the study of how punishment (among other) strategies by trustors and/or trustees might influence the evolution of trust in spatial structures. We hope our work will provide a perspective for further exploring classic two-role trust games.

\section*{Acknowledgement}
A.S. was supported by the National Research, Development and Innovation Office (NKFIH)
under Grant K142948.

\section*{Data Availability Statement}
The data that support the findings of this study are available within the article.

\appendix
\section{Well-mixed populations}\label{sec_wellmix}
Here, we analyze the model in an infinite and well-mixed population. The frequencies of Investing/Non-investing trustors and Trustworthy/Untrustworthy trustees are still denoted by $\rho_\text{I}$, $\rho_\text{N}=1-\rho_\text{I}$, $\rho_\text{T}$, and $\rho_\text{U}=1-\rho_\text{T}$, respectively. 

By the payoff calculation in Eq.~(\ref{eq_m1}), we have a trustor's expected payoff $\Bar{\pi}_\text{I}$ for strategy I and $\Bar{\pi}_\text{N}$ for strategy N,
\begin{subequations}
    \begin{align}
        \Bar{\pi}_\text{I}&=
        \rho_\text{T} (1-x+urx)+(1-\rho_\text{T})(1-x), \\
        \Bar{\pi}_\text{N}&=
        1.
    \end{align}
\end{subequations}
Similarly, according to the payoff calculation in Eq.~(\ref{eq_m2}), we have a trustee's expected payoff $\Bar{\pi}_\text{T}$ for strategy T and $\Bar{\pi}_\text{U}$ for strategy U,
\begin{subequations}
    \begin{align}
        \Bar{\pi}_\text{T}&=
        \rho_\text{I} u(1-r)x, \\
        \Bar{\pi}_\text{U}&=
        \rho_\text{I} ux.
    \end{align}
\end{subequations}

According to replicator dynamics~\cite{taylor1978evolutionary}, the evolution of $\rho_\text{I}$ and $\rho_\text{T}$ can be expressed as
\begin{subequations}
    \begin{align}
        \dot{\rho}_\text{I}&=\rho_\text{I} (1-\rho_\text{I})(\Bar{\pi}_\text{I}-\Bar{\pi}_\text{N})=\rho_\text{I} (1-\rho_\text{I})(\rho_\text{T} ur-1)x, \\
        \dot{\rho}_\text{T}&=\rho_\text{T}(1-\rho_\text{T})(\Bar{\pi}_\text{T}-\Bar{\pi}_\text{U})=-\rho_\text{T}(1-\rho_\text{T})\rho_\text{I} urx.
    \end{align}
\end{subequations}
It follows that $\dot{\rho}_\text{T}\leq 0$ always holds. When $\rho_\text{I}=0$, the value of $\rho_\text{T}$ is in neutral drift. When $\rho_\text{I}>0$, the value of $\rho_\text{T}$ has a stable point $\rho_\text{T}=0$, and as a result, $\dot{\rho}_\text{I}< 0$ always holds and the stable point $\rho_\text{I}=0$ for trust extinction is unavoidable.

%

\end{document}